\documentclass[a4paper,fleqn,usenatbib]{mnras}
\usepackage{amssymb,amsmath,graphicx,psfrag,mathtools}
\usepackage[T1]{fontenc} 
\usepackage{ae,aecompl} 
\usepackage{url}
\title[Lunar-Scattered Galactic Micro-FRB]{Searching for Galactic Micro-FRB
with Lunar Scattering}
\author[J. I. Katz]{
J. I. Katz,$^{1}$\thanks{E-mail katz@wuphys.wustl.edu} 
\\
$^{1}$Department of Physics and McDonnell Center for the Space Sciences,
Washington University, St. Louis, Mo. 63130 USA 
}
\date{Accepted XXX.  Received YYY; in original form ZZZ} 
\pubyear{2020} 
\date{\today}
\begin{document} 
\psfrag{Reflectivity X cos(theta)}{Reflectivity$\times \cos{\theta}$}
\psfrag{Angle of Incidence (theta)}{Angle of incidence ($\theta$)}
\label{firstpage} 
\pagerange{\pageref{firstpage}--\pageref{lastpage}} 
\maketitle 
\begin{abstract}
	Does the Galaxy contain sources of micro-FRB, {lower energy
	events resembling the known FRB but detectable only at Galactic
	distances}?  The answer to this question is essential to determining
	the nature of FRB sources.  At typical ($10$ kpc) Galactic distances
	a burst would be about 117 dB brighter than at a ``cosmological''
	($z = 1$) distance.  The radiation of Galactic micro-FRB, if they
	exist, could be detected after Lunar reflection, or an upper bound
	on their rate set, by a modest (20 m at 1.4 GHz) radio telescope
	staring at the Moon.  It would have all-sky sensitivity.  The delay
	between detection of direct {(by STARE2 or dipoles)} and
	Lunar-scattered radiation would restrict a burst's position to a
	narrow arc.
\end{abstract}
\begin{keywords} 
radio continuum: transients, Galaxy: general, instrumentation: miscellaneous
\end{keywords} 
\section{Introduction}
\label{intro}
The sources of Fast Radio Bursts (FRB) remain a mystery.  {Many
different astronomical objects and processes have been proposed
\citep{Theory}.}  If associated with stars or their remnants {(neutron
stars, stellar mass black holes, supernova remnants {\it etc.\/})}, the
distribution of FRB fluence on the sky would be expected to {resemble
that of Galactic stars,} concentrated in the Galactic plane, as is
the distribution of other radiation associated with {stars and their
remnants} \citep{K19a}\footnote{Observed gamma-ray bursts (GRB) are
isotropically distributed, but, integrated long enough ($\gg 10^6$ y) to
include expected infrequent Galactic GRB, the distribution of GRB fluence is
expected also to be concentrated in the Galactic plane.}.  This is a
consequence of the domination of the distribution of stellar mass in the
Universe, weighted by the $-2$ power of its distance, by the Galactic disc.

Yet FRB are isotropically distributed \citep{B18}.  Are FRB associated with
stars or with some unrelated class of objects?  If the former, concentration
of micro-FRB, such as might be produced by repeating FRB, in the Galactic
plane would be expected.  Empirical confirmation or contradiction of that
prediction would help decide the question of the origin of FRB.

Some FRB repeat,
requiring a non-catastrophic origin, but it is not known if apparently
non-repeating FRB actually repeat at long intervals or are the results of
catastrophic non-repeating processes.  Phenomenological arguments for the
distinct nature of the sources of repeating and apparently non-repeating
FRB have been recently presented by \citet{K19a,LZNS19}, while \citet{R19}
has argued on statistical grounds that apparent non-repeaters must actually
repeat because the rates of known catastrophic events are insufficient.

{Here I consider the possibility of detecting micro-FRB, events
intermediate in energy between giant pulsar pulses and FRB observed at $z
\sim 1$.  Even if micro-FRE were 12 orders of magnitude less energetic than
``cosmological'' FRB they would be observable at Galactic ($\sim 10$ kpc)
distances.} There may be two contributors to a population of Galactic
micro-FRB: Repeaters that, like FRB 121102, have large numbers of weak
bursts but that are much weaker or less active than FRB 121102 (which, if at
Galactic distances, would have been detected in side-lobes of unrelated
radio observations), and possible weak non-repeaters.

It is known \citep{K19} that individual FRB sources produce bursts with a
wide range of strengths, suggesting that even weak or comparatively inactive
sources may produce bursts detectable at Galactic distances.  If FRB are
produced by comparatively common objects like neutron stars, of which there
are many in the Galaxy with a broad range of parameters, then a minority
with optimal parameters (such as very young neutron stars) may be detectable
at cosmological distances while the much greater number with less favorable
parameters might produce micro-FRB detectable only at Galactic distances.

Most natural events that leave their sources fundamentally unchanged repeat,
with a spectrum of outburst sizes that increases rapidly towards weaker
outbursts.  Examples include earthquakes, Solar and stellar flares, giant
pulsar pulses, lightning and SGR (soft gamma repeater) outbursts.  Most of
these processes appear to have no natural size scale, but rather a power law
distribution of event sizes.  In contrast, the largest SGR outbursts and
recurrent nov\ae\ appear to be exceptions, with characteristic sizes.
Catastrophic events that destroy their sources, such as supernov\ae\ and
gamma-ray bursts, generally also have characteristic sizes.

FRB energetic enough to be observed by Parkes at cosmological distances
occur at a rate $\sim 10^6$/sky-y \citep{CC19}.  With $\sim 10^9$ $L^*$
galaxies with $z \le 1$ \citep{CWDM16} their rate is $\sim
10^{-3}$/galaxy-y.  Less energetic bursts of repeating FRB, detectable at
Galactic distances, may be frequent enough to occur in feasible observing
times \citep{B20}.  Some constraints on those FRB, with energies between
those detectable only at Galactic distances and those detectable at
cosmological distances, were set by observations of the Virgo cluster
\citep{A19}.  Detection of Galactic micro-FRB would establish that the
Galaxy contains many sources, consistent with the popular hypotheses
\citep{K12,CW16,K18,CC19} that these sources are neutron stars.  The absence
of Galactic micro-FRB would point to sources rare enough that there are {\it
none} in the Galaxy \citep{K19a,K19b}, likely excluding neutron star origin.

In this paper I suggest a method of monitoring the entire Galaxy for
micro-FRB.  Its sensitivity would be about 73 dB less than that of pointed
observations with a Parkes-class telescope, but this is more than
compensated by the $\approx$ 117 dB inverse-square law ratio of intensity of
Galactic sources in comparison to those at ``cosmological'' distances ($z =
1$; luminosity distance 7 Gpc).  The FRB K-correction is unknown but the
redshift also tends to make Galactic counterparts brighter than distant FRB.
The survey would probe the FRB luminosity function about 44 dB deeper than
is possible at cosmological distances.  The proposed observations would be
sensitive to bursts over almost the entire sky, and the hypothesis of a
numerous population of Galactic micro-FRB could be confirmed or excluded.

The Moon reflects, mostly as a specular glint but partly diffusely, radio
radiation that illuminates it.  Radiation from a FRB in any direction
(except for the narrow cone eclipsed by the Earth) is reflected in every
direction and can be detected at the Earth (except for sources in the narrow
cone eclipsed at the Earth by the Moon).  The intensity at the Earth is much
less than the incident intensity at the Moon, but the great brightness of a
Galactic FRB, in comparison to the same event at cosmological distances,
more than compensates for this.

For a radio telescope beam matched to the angular size of the Moon the loss
in sensitivity, aside from a contribution $\sim 13$ dB attributable to the
Lunar reflectivity and a trigonometric factor, equals the gain in acceptance
solid angle ($4\pi$ sterad {\it vs.\/} the beam solid angle). In addition,
the required telescope would be modest ($\approx 20$ m in L band) and
perhaps possible to dedicate to these observations whenever the Moon is 
above the horizon, thousands of hours per year.  A larger telescope (or one
observing at higher frequency) would have greater sensitivity, at the price
of requiring multiple beams to cover the Moon.

{The proposed observations might be combined with other radio astronomy
programs \citep{M18,J19} that involve staring at the Moon.}
\section{Lunar Scattering}
Scattered radiation can be detected by a radio telescope staring at the
Moon.  The glint of a spherical Moon is not significantly spread in time
(diffraction broadens it very slightly from a geometrical glint).

Most direct measurements \citep{TD66} of Lunar radio-wave
reflection have been monostatic and are not directly applicable to the
bistatic problem of scattering towards the Earth from general directions of
incidence.  The measured electromagnetic properties \citep{OS75} of the
Lunar surface can be used to estimate the bistatic scattering.  At the
L-band frequencies of most FRB observations the properties in the upper
$c/n\omega = 1/k = \lambda/2\pi \sim$ 1--3 cm of soil are applicable: the
density is about 1.55 g/cm$^3$ and the empirical relation of \cite{OS75}
indicates a dielectric constant $K^\prime \approx 2.77$ and refractive index
$n = \sqrt{K^\prime} \approx 1.66$.

\begin{figure}
	\centering
	\includegraphics[width=3.3in]{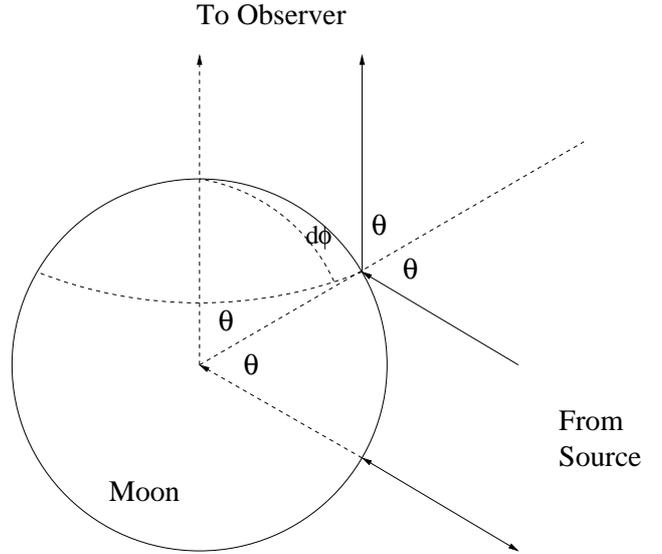}
	\caption{\label{geom}Specular reflection of radiation from a source
	at infinity towards the Earth, also approximated as at infinity, by
	a spherical Moon.}
\end{figure}

Here we treat the Moon as a specularly reflecting sphere of radius $R$ at a
distance $D \gg R$ from the observer.  For an observer on the Earth $D/R
\approx 220$ and $D/R_E \approx 60$ so that both source and observer may be
considered to be at infinity.  The geometry is shown in Fig.~\ref{geom}.

On scales $\gtrsim 20$ m median lunar slopes are $\sim 5\text{--}10^\circ$
\citep{SBA95,R11}).  {The light travel time across a Lunar radius is 6
ms.  Smooth slopes contributing glints will be separated by a small fraction
of the Lunar radius, possibly contributing to a temporal width of the glint
$\sim 1$ ms.  Diffusely scattered radiation may arrive with delays of up to
$12\cos{\theta}$ ms, contributing a weak tail to the specularly scattered
pulse (Fermat's Principle ensures that the specular glint arrives first).}

If the FRB signal is $F$ per unit area on a surface perpendicular to its
direction, it delivers $F \cos{\theta}$ per unit area of the surface of the
Moon at the specular point, where $\theta$ is the angle between the specular
point and the direction to the Earth (Fig.~\ref{geom}).  The signal may be
the flux, the fluence or the integral of the product of flux with an
arbitrary function of time.  The signal illuminating the portion of the
Lunar surface in an interval $d\theta$ about the angle of incidence $\theta$
and $d\phi$ from the plane of incidence ($\phi$ is the azimuthal angle
around the direction to the Earth) $dF = F \cos{\theta} R^2 \sin{\theta}
d\theta d\phi$.  A square patch is defined by $d\phi = \csc{\theta}
d\theta$.  This is reflected into a solid angle $d\Omega = \sin{\theta}\,
2d\theta\,2d\phi = 4 (d\theta)^2$ because the change in direction on
reflection is twice the angle of incidence.

\begin{figure}
	\centering
	\includegraphics[width=3.3in]{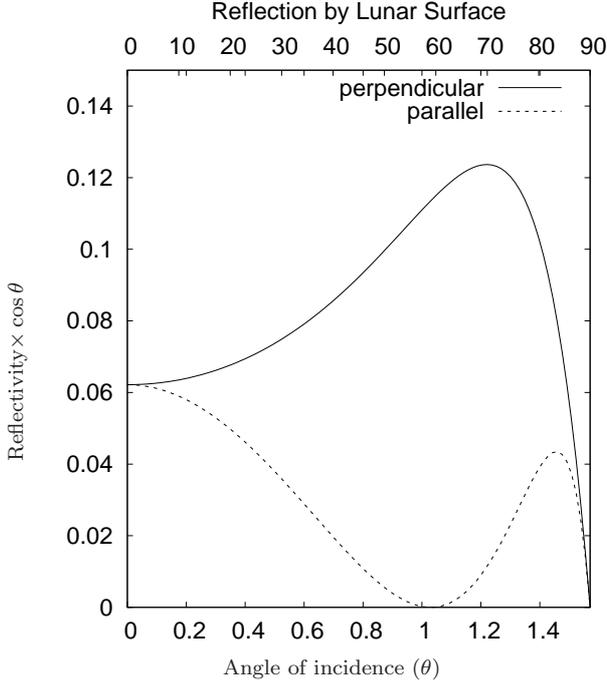}
	\caption{\label{refl}The angle-dependent factors in Eq.~\ref{obs}
	of the two polarization states for a Lunar dielectric constant
	$K^\prime = 2.77$ ($n = 1.66$).  Lower $x$-axis gives angle of
	incidence $\theta$ in radians, upper in degrees.}
\end{figure}

The reflected signal per unit solid angle
\begin{equation}
	{dF \over d\Omega} = {1 \over 4} F R^2 \cos{\theta}.
\end{equation}
At a distance $D$ the observed signal is
\begin{equation}
\label{obs}
	F_\mathrm{obs} = F {{\cal R(\theta)} \over 4} {R^2 \over D^2}
	\cos{\theta},
\end{equation}
where $\cal R$ is the reflection coefficient.  For polarizations
perpendicular and parallel to the plane of incidence
\begin{equation}
	\label{calR}
	\begin{split}
	{\cal R}_\perp(\theta) &= \left({\cos{\theta} -
	\sqrt{n^2 - \sin^2{\theta}} \over \cos{\theta} +
		\sqrt{n^2 - \sin^2{\theta}}}\right)^2;\\
	{\cal R}_\parallel(\theta) &= \left({n^2 \cos{\theta} -
		\sqrt{n^2 - \sin^2{\theta}} \over
		n^2 \cos{\theta} + \sqrt{n^2 - \sin^2{\theta}}}\right)^2.
	\end{split}
\end{equation}
The products ${\cal R}(\theta) \cos{\theta}$ are shown in Fig.~\ref{refl}.

There are two important consequences of Eqs.~\ref{obs} and \ref{calR} and
Fig.~\ref{refl}:
\begin{enumerate}
	\item As a rough approximation
		\begin{equation}
			\label{grefl}
			{F_\mathrm{obs} \over F} \sim 0.02 {R^2 \over D^2}
			\sim 4 \times 10^{-7};
		\end{equation}
		the scattered signal is about 64 dB weaker than the direct
		signal\footnote{There is a selection bias towards detection
		at angles at which this factor is larger, so we take a
		factor somewhat larger than the mean of Fig.~\ref{refl}.}.
	\item Scattering is strongly polarizing.
\end{enumerate}

The optimal diameter of a single-beam telescope is $d_\mathrm{tel} \approx
\lambda D/(2R) \approx 22$ m at 1400 MHz.  Many such telescopes exist, and
would provide nearly $4\pi$ sky coverage for FRB during the 0.5 duty factor
of the Moon above the horizon.  The telescope gain near the center of its
beam
\begin{equation}
	\label{gtel}
	g_\mathrm{tel} \approx 10 \log_{10}{\left({\pi d_\mathrm{tel} \over
	\lambda}\right)^2} \approx 10 \log_{10}{\left({\pi D \over 2 R}
	\right)^2} \approx 51\,\text{dB}.
\end{equation}
Combining Eqs.~\ref{grefl} and \ref{gtel} leads to a system gain
\begin{equation}
	\label{gsys}
	g_\mathrm{sys} \approx 10 \log_{10}{\left[0.02\left({\pi \over 2}
	\right)^2\right]} \approx -13\,\text{dB},
\end{equation}
independent of $D/R$; the weakness of Lunar-scattered radiation is
cancelled, except for the reflectivity and a geometric factor, by the gain of
a telescope beam matched to the angular size of the Moon.

The gain of the proposed system should be compared to that of a
single dipole, which is near 0 dB (except in its narrow nulls), and to that
of the Parkes telescope ($d_\mathrm{tel} = 64$ m) in the center of one of
its beams, which is about 60 dB.  The proposed system would be about 13 dB
less sensitive than a single dipole and 73 dB less sensitive than Parkes,
corresponding to a detection threshold $\sim 20$ MJy for a $\sim 1$ ms
burst.  In a search for Galactic micro-FRB this lesser sensitivity, in
comparison to Parkes, would be compensated by the $4\pi$ acceptance angle
and by the fact that at Galactic distances (10 kpc) a given burst
would be about 117 dB brighter than at luminosity distance 6.7 Gpc ($z = 1$)
and about 100 dB brighter than at luminosity distance 1.0 Gpc ($z=0.193$,
the distance of FRB 121102).  The proposed system would probe the FRB
luminosity function to energies $117 - 73 = 44$ dB less than Parkes at $z =
1$ and to energies $100 - 73 = 27$ dB less than Parkes at $z = 0.193$.
\section{Localization}
If a burst were detected by the proposed system,
localization would be necessary to confirm the burst's  Galactic nature and
to permit further investigation.  There are two methods of localization,
both of which require simultaneous detection with another instrument.

One such instrument could be a dipole or array of dipoles.  Even a single
dipole would be more sensitive than the proposed system, but would suffer
from a high rate of electromagnetic interference because of a dipole's
roughly isotropic sensitivity.  A dipole would detect a Galactic FRB with
the same energy as the Parkes FRB at $z \sim 1$ with signal-to-noise ratio
roughly 50 dB higher than the detections of the cosmological FRB
\citep{K14,M17}.  A phased array of dipoles would provide higher sensitivity
over the entire visible hemisphere, and also directional information to
discriminate against interference, whether through the antenna or
``back-door'' into the electronics.  STARE2 \citep{B20} has somewhat lower
angular acceptance (3.6 sterad) than dipoles and a sensitivity of 300 kJy
for 1 ms bursts, roughly 55 dB less than that of Parkes but about 5 dB
better than that of a dipole.
\subsection{Temporal}
Comparing the arrival times of a burst at a Lunar-staring telescope and at
another receiver, such as a dipole or an array of dipoles, would constrain
the position of the burst.  Radiation reflected by the Moon arrives later
than that observed directly by
\begin{equation}
	\Delta t = {D \over c} \left(1 - \cos{(\pi - 2\theta)}\right) = 
	{D \over c} \left(1 + \cos{(2\theta)}\right) \sim 1\,\text{s},
\end{equation}
where $D$ is the distance to the Moon.
Bursts are typically 1--10 ms long but contain temporal structure as fine as
$\delta t \sim 30\,\mu$s \citep{CC19}.  On the basis of the time difference
of arrival between the Lunar-reflected signal and the direct signal detected
by a dipole array or STARE2, a burst could be localized to an arc of width 
\begin{equation}
	\label{dtheta}
	\Delta \theta \sim {\delta t \over |d\Delta t/d\theta|} \sim
	5^{\prime\prime},
\end{equation}
where $\delta t \sim 30\,\mu$s is the uncertainty in the time difference.
If phase coherence were maintained between the detectors, $\delta t$ in
Eq.~\ref{dtheta} would be replaced by $\lambda/(2 \pi c)$.  This narrow,
albeit one-dimensional, localization of bursts from anywhere on the sky is
the chief advantage of observation of Lunar-scattered burst radiation.

Less sharp localization $\delta \theta \sim c\delta t/L \sim 5
(\text{6000 km}/L)^\prime$ transverse to the narrow arc can be provided by
comparing arrival times at two dipole arrays separated by a distance $L$.
Very large separations reduce the probability that a burst is above the
horizon at both locations, so $\sim 6000$ km may be a practical upper limit
on useful values of $L$.
\subsection{Polarization}
Fig.~\ref{refl} shows that the polarization of reflected radiation depends
on its angle of incidence.  Combining the measured polarization of a
Lunar-reflected burst with that measured directly by a dipole or dipole
array would constrain the direction of origin of the burst.  Because of
the limited accuracy of polarization measurements this could not be a tight
constraint, probably no narrower than $\sim 0.2$ rad, but its intersection
with a temporal arc of localization would be sufficient to establish whether
a burst originated within, or outside, the Galactic plane.
\subsection{No Solar Reflection}
If reflection from the Sun could be observed, it would provide an
independent and precise second temporal localization arc intersecting that
of Lunar reflection.  Unfortunately, the Solar reflectivity at frequencies
at which FRB are observed is extremely small.  

The free-free absorption coefficient of plasma \citep{S62}
\begin{equation}
	\begin{split}
	\kappa &= {4 \over 3}\sqrt{2 \pi \over 3} {Z^2 e^6 n_e n_i \over c
	m_e^{3/2} \nu^2 (k_B T)^{3/2}} g_{ff}\\ &= \kappa_0
	\left({\text{1400 MHz} \over \nu}\right)^2
	\left({T \over \text{25000 K}}\right)^{-3/2} n^2\ \text{cm}^{-1},
	\end{split}
\end{equation}
where $\kappa_0 \approx 9.3 \times 10^{-27}$ cm$^5$ and we have taken a
singly ionized plasma with $n = n_e = n_i$ and $T = 25000$ K (the
temperature at the turning point where the plasma frequency $\nu_p =
\text{1400 MHz}$).  The critical electron density
\begin{equation}
	n_c = {\pi m_e \nu^2 \over e^2} = 2.4 \times 10^{10}
	\left({\nu \over \text{1400 MHz}}\right)^2 \text{cm}^{-3}.
\end{equation}
At 1/2 scale height above the turning point the Gaunt factor $g_{ff} \approx
4.1$ (the wave group velocity is 0.63 c).  The density distribution
\begin{equation}
	n(z) = n_c \exp{(-z/h)},
\end{equation}
where the scale height
\begin{equation}
	h = {k_B T R_\odot^2 \over G M_\odot \mu} \approx 1.17 \times 10^7
	\left({T \over \text{25000 K}}\right) \text{cm},
\end{equation}
taking the mean molecular weight $\mu = 1.29 m_p/2$ of singly ionized Solar
plasma.  The integrated round trip optical depth on a radial ray from
infinity to the critical density at which the wave reflects
\begin{equation}
	\begin{split}
	\tau &= 2 \kappa_0 n_c^2 \int_0^\infty\!dz\,\exp{(-2z/h)}\\
	&\approx 640 \left({\nu \over \text{1400 MHz}}\right)^2
	\left({T \over \text{25000 K}}\right)^{-1/2};
	\end{split}
\end{equation}
the absorption on non-radial rays requires a numerical integral but is also
large.

Lower frequencies are reflected higher in the Solar atmosphere where
temperatures are higher, but the dependence on temperature is not steep.
Solar reflection is likely not observable for $\nu \gtrsim 100$ MHz, several
times lower than the lowest frequencies at which FRB have been observed.
\section{Rates}
{ If FRB-like sources are distributed in all galaxies, including our own,
the detectable event rate is related to the distribution of FRB energies.}
Extrapolation of the rate of FRB observed at $z \lesssim 1$ by a power law
with exponent $-\alpha$ would predict an all-sky detection rate of {
Galactic micro-FRB} $\sim 10^{-3} \times (F_\mathrm{cosmo}/
F_\mathrm{Lunar})^\alpha$/y, where $F_\mathrm{cosmo}/ F_\mathrm{Lunar}
\approx 2.5 \times 10^4$ ($117 - 73 = 44$ dB) is the ratio of detectable
{ emitted powers or energies and $10^{-3}$/y is the FRB rate per galaxy
estimated from Parkes observations {(Sec.~\ref{intro})}.  The rate of
detectable Galactic micro-FRB is $> 1$/y if $\alpha > 0.7$}.

{ The proposed system would be about 73 dB less sensitive than Parkes,
implying an equivalent detection threshold (the flux density impinging on
the Moon, not on the receiver) of about 20 MJy for a 1 ms burst, about 70
times less sensitive than STARE2 \citep{B20}.  For STARE2 the extrapolated
burst detection rate is $\sim 10^{-3} (F_\mathrm{cosmo}/
F_\mathrm{STARE2})^\alpha$/y, where $F_\mathrm{cosmo}/F_\mathrm{STARE2}
\approx 1.7 \times 10^6$ ($117 - 55 = 62$ dB).  The observed upper bound of
40/y implies $\alpha < \log_{10}{(40/10^{-3})}/6.2 \approx 0.7$, suggesting
a detection rate for the Lunar reflection system $\lesssim 1$/y, but the
validity of assuming and extrapolating a power law over several orders of
magnitude is questionable.

Direct comparison, depending only on the assumed power law slope between the
detection thresholds of the Lunar and STARE2 systems indicates that the
Lunar system would have a detection rate $\sim (4 \pi/3.6) 70^{-\alpha}$
times that of STARE2, where the first factor is the ratio of their
acceptance solid angles when the Moon is above the horizon and the second
factor is the ratio of their sensitivities.  For $\alpha = 0.7$ this is
about 0.2 if the Moon is continuously observed, requiring at least two
widely separated telescopes; with only one the Moon is below the horizon
half the time and the factor becomes about 0.1.}
\section{Discussion}
Do micro-FRB exist in our Galaxy at all?  The answer has implications for
the nature of FRB sources \citep{K12,K19a,K19b}.  {Proposed
non-catastrophic FRB models \citep{Theory} do not make makes quantitative
predictions of the numbers of possible micro-FRB.}  Detection of even one
would establish that their sources are comparatively common in the stellar
population.  Non-detection would tend to exclude sources, like neutron
stars, that are present in the Galaxy, {but this conclusion would depend on
the extrapolation of the observed FRB rate to low energy bursts.  In the
extreme limit (inconsistent with observations of repeating FRB) that bursts
are standard candles, no inference could be drawn unless no bursts were
observed for a time several times the mean recurrence time per galaxy of
known cosmological FRB, $\gg 10^3$y.}

Non-repeating FRB may be produced with catastrophic events with a natural
scale (so that at low energy $\alpha < 0$).  Recurring phenomena generally
have more weak events than strong ones, typically with a power law
distribution that grows rapidly towards small amplitude ($\alpha > 0$).
This appears to be true for FRB 121102, the only well-studied repeater,
although this has not been well quantified and extrapolation to weaker
bursts detectable only at Galactic distances is speculative.  Despite this,
the existence of repeating FRB with a broad distribution of intrinsic
strengths suggests that if any such sources are present in the Galaxy their
weaker bursts might be observable.

{Observation of FRB radiation reflected by the Moon:
\vspace{-\topsep}
\begin{enumerate}
	\item simultaneously observes nearly the entire sky when the Moon
		is above the horizon;
	\item requires the use of only a single radio telescope, of $\sim
		22$ m diameter for observations at $\sim 1$ GHz;
	\item in combination with an array of dipoles or STARE2 observing
		the direct radiation, can localize sources to a narrow
		arc on the sky.  Approximate localization by two or more
		STARE2 receivers or arrays of dipoles can restrict the
		location to a small portion of that arc.
\end{enumerate}
The proposed method is not intrinsically more sensitive to the direct
radiation of a FRB than the far sidelobes of a large telescope, which
\cite{T16} used to demonstrate the absence of a FRB associated with the
giant December 27, 2004 outburst of SGR 1806$-$20, but lunar scattering
makes localization possible.}

\section*{Acknowledgement}
I thank C. D. Bochenek and L. Cowie for useful discussions.

\bsp 
\label{lastpage} 
\end{document}